# Measurement of Differential Static Polarizability and Frequency of an Inner-Shell Orbital Clock Transition in Lattice-Trapped $^{174}$Yb

Yuan Yao, Congyu Wang, Jinpeng Zou, Haosen Shi, Yuqing Xin, Longsheng Ma, and Yanyi Jiang[*]

State Key Laboratory of Precision Spectroscopy, East China Normal University, Shanghai 200241, China

Additional clock transitions of ytterbium atoms based on inner-shell orbital transition could benefit the search for new physics beyond the Standard Model. Observation of these transitions with high resolution is a prerequisite for making precise frequency measurements. Here, we observe 4.3 Hz-linewidth spectra of the inner-shell orbital transition at 431 nm in lattice-trapped $^{174}$Yb. With high-resolution spectra, we precisely determine the differential static polarizability of the transition to be −2.10(4) kHz/(kV/cm)$^2$. The magnitude of this polarizability is approximately 1/17 of that of the well-known clock transition in $^{171}$Yb at 578 nm, indicating a reduced sensitivity to blackbody radiation. We carry out a frequency ratio measurement between the two clock transitions of ytterbium atoms with an uncertainty of 9×10$^{−15}$. The frequency of the 431 nm transition is determined to be 695 175 030 801 776.5(6.3) Hz. These results represent a step forward in future studies on the search for new physics beyond the Standard Model.

## I. INTRODUCTION

Optical atomic clocks have demonstrated unprecedented fractional frequency uncertainties and instabilities at the 10$^{−18}$ level or even lower [1–5]. Such precise and accurate clocks are candidates for the next-generation primary frequency standard [6]. Meanwhile, they are indispensable tools in fundamental research [1, 7–10]. The search for new physics beyond the Standard Model through low-energy experiments now relies on extremely accurate optical atomic clocks. In such applications, at least two clock transitions with different sensitivities to new physics are employed to accurately measure possible tiny frequency changes. When two clock transitions are detected in the same apparatus, whether involving different atom species or a single species, the system will enjoy the greatest common-noise rejection [7, 11]. A good example is the test of temporal variation of the fine structure constant and the search for violations of the Einstein equivalence principle, which are conducted using the electric-quadrupole transition and the electric-octupole transition of Yb$^+$ [7].

Recently, an inner-shell orbital clock transition of neutral ytterbium (Yb) atoms has been proposed [12, 13]. This transition occurs from the ground state $4f^{14}6s^2$ $^1S_0$ to the long-lived excited state $4f^{13}5d6s^2$ ($J = 2$). According to calculations, this new transition at a wavelength of 431 nm has a natural transition linewidth that is nearly ten times narrower and a blackbody radiation (BBR) shift that is seven times lower than those of the well-studied $^1S_0$–$^3P_0$ clock transition at 578 nm [12, 13]. This indicates its potential to further reduce systematic uncertainties. Moreover, it exhibits high sensitivity to variations of the fine-structure constant. Dual clock transition operation, which involves both the 578 nm and 431 nm transition of neutral Yb atoms, promises an even more precise test of fundamental physics, owning to its potentially lower frequency instability enabled by the large number of neutral atoms involved.

After the proposal of using the inner-shell orbital clock transition of neutral Yb atoms, Ishiyama *et al*. first observed the transition with a 30 kHz linewidth for all isotopes in a crossed far-off resonance trap [14]. They also observed a 12 kHz-linewidth spectrum in lattice-trapped $^{174}$Yb atoms. Recently, they observed a spectrum with an 80 Hz linewidth in a three-dimensional optical lattice trapped $^{174}$Yb atoms [15]. The absolute frequency of the $^1S_0$ ($F = 1/2$) –$4f^{13}5d6s^2$ ($J = 2, F = 3/2$) transition of $^{171}$Yb was measured by two groups, and the results agreed with each other within the measurement uncertainty at the kHz level [16, 17]. Qiao *et al*. also observed a 4 kHz-linewidth spectrum of $^{171}$Yb atoms, but the transition was excited from the metastable state of $^3P_0$ to the $4f^{13}5d6s^2$ ($J = 2$) state at a wavelength of 1695 nm. They stabilized the clock laser to the transition to realize a closed-loop operation of an optical clock with a fractional frequency instability of $10^{-12}/\sqrt{\tau}$, where $\tau$ is the averaging time [18]. To date, the spectral linewidth of this transition has not yet reached the hertz level, limiting both the measurement precision and accuracy. However, tests of fundamental physics related to this transition are being conducted in parallel [15, 19].

In this Letter, we first observe the inner-shell orbital clock transition of $^{174}$Yb at 431 nm with a spectral linewidth as narrow as 4.3 Hz, an improvement of more than an order of magnitude compared to previous work [15]. Such a spectral linewidth is comparable to that of state-of-the-art optical atomic clocks. Based on 8 Hz linewidth transition spectra, we stabilize a 431 nm laser to the clock transition. Through interleaved measurement, we determine the differential static polarizability of the 431 nm transition to be $-2.10(4)$ kHz/(kV/cm)$^2$. The magnitude of this polarizability is nearly 1/17 of that of the 578 nm clock transition [20]. Such a small differential static polarizability promises a much lower sensitivity to the BBR shift, and thus a smaller systematic uncertainty since the BBR shift is the main limitation for most state-of-the-art optical clocks based on neutral atoms [1, 4]. Using an optical frequency divider, we measure the frequency ratio between the 431 nm inner-shell orbital clock transition and the 578 nm clock transition. After correcting frequency shifts, we determine the frequency of the $^1S_0$ (m$_F$ = 0) $-4f^{13}5d6s^2$ ($J$ = 2, m$_F$ = 0) transition to be 695 175 030 801 776.5 Hz with an uncertainty of 6.3 Hz, which is dominated by lattice shift and statistical uncertainty.

## II. SUB-10 HZ-LINEWIDTH RABI SPECTRA

After a two-stage magneto-optic trap, up to 3000 $^{174}$Yb atoms are loaded into a one-dimensional optical lattice at a wavelength of 798 nm, which is generated from a continuous wave Ti:Sapphire laser [17]. As shown in Fig. 1(a), the optical lattice is tilted from the vertical axis by about 5°. The operating trap depth of the lattice is 100 $E_r$ with $E_r = (h\nu_{at})^2/(2m_{Yb} \times c^2)$, where $h$ *is* the Planck constant, $\nu_{at}$ is the lattice frequency, $m_{Yb}$ is the mass of $^{174}$Yb atom, and $c$ is the speed of light. The diameter of the optical lattice is about 110 μm. Laser light at 431 nm is aligned along the lattice light to probe the $^{174}$Yb atoms trapped in the optical lattice. Via an optical frequency divider [17, 21], the frequency of the 431 nm laser light is referenced to a cavity-stabilized 578 nm laser, which is further stabilized to the 578 nm clock transition of $^{171}$Yb. The relationship is given by $v_{431-L} = v_{578}/R$, where $R$ is a preset divisor. We use phase-noise compensated optical fiber links to transfer the 431 nm laser light to the apparatus of the lattice-trapped Yb atoms. A magnetic field $\vec{B}_{bias}$ (quantization axis) with a magnitude of 1.5 G is applied at an angle $\theta$ with respect to the polarization of the lattice light $\vec{e}_{lat}$ elat and an angle of $\phi$ with respect to the propagation direction of the probe light $\vec{k}_{431}$ (see Sec. S1 in Supplemental Material).

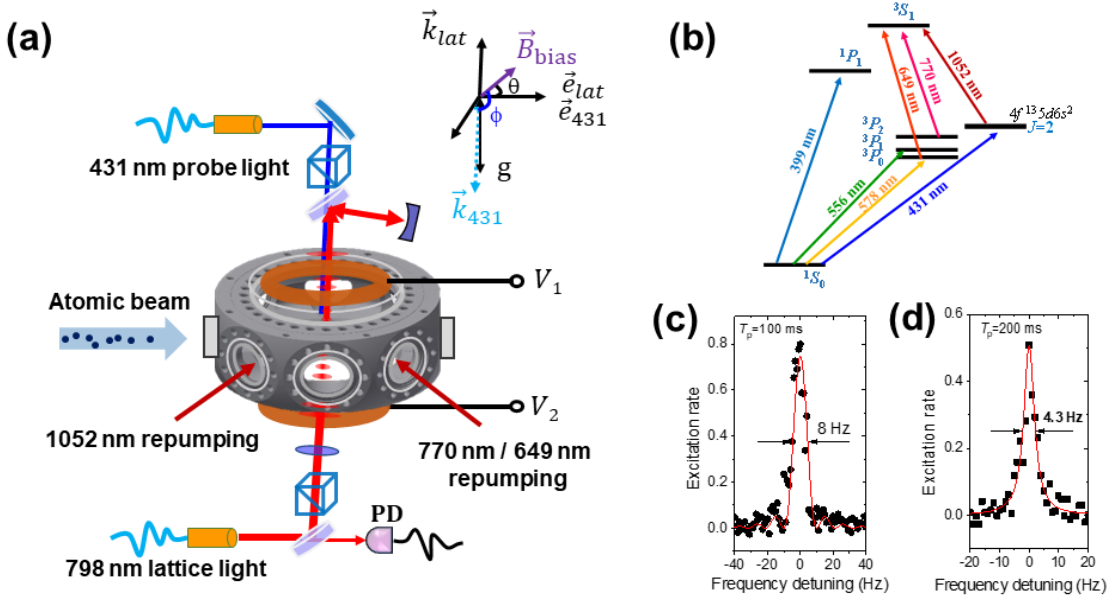

FIG. 1. Experimental setup for precision spectroscopy. (a) Schematic diagram of probing inner-shell orbital clock transition of lattice-trapped $^{174}$Yb atoms. (b) Energy levels relevant to our experiments. (c) and (d) Rabi spectrum (no averaging) when the 431 nm probe laser light is on for 200 ms and 100 ms, respectively. The red solid lines are the fit to the data.

To minimize Zeeman frequency shift, we perform a Rabi spectroscopy of the $^1S_0$ ($m_F = 0$) – $4f^{13}5d6s^2$ ($J = 2$, $m_F = 0$) transition. The 431 nm probe light with a power of 45 μW and a diameter of approximately 1 mm is used to excite the transition. The polarization of the 431 nm probe light, $\vec{e}_{431}$, is aligned along $\vec{e}_{lat}$. In each frequency sweep step, the populations in the ground and the excited states are separately measured using the shelving detection technique, in which the atoms in the excited state are repumped back to the ground state via the channels of $4f^{13}5d6s^2$ ($J = 2$) → $^3S_1$ (1052 nm), $^3P_0$ → $^3S_1$ (649 nm), and $^3P_2$ → $^3S_1$ (770 nm), as shown in Fig. 1(b) for relevant energy levels.

The total differential polarizability between the ground state and the excited state at the lattice angular frequency $\omega_l$ is

$$\Delta\alpha(\omega_l) = \Delta\alpha^S(\omega_l) - \frac{3\cos^2\theta - 1}{2}\Delta\alpha^T(\omega_l), \tag{1}$$

where $\Delta\alpha^S$ and $\Delta\alpha^T$ denotes differential scalar and tensor polarizability, respectively [13, 14]. Here, we set $\theta \approx 54.8°$ to make the second term on the right side of Eq. (1) equal to zero (see Sec. S2 in Supplemental Material). Meanwhile, we tune the lattice frequency to satisfy $\Delta\alpha(\omega_l)=0$. Thereby, we can achieve a narrow linewidth spectrum due to a negligible Stark shift experienced by all the atoms scattered in lattice sites. When either the magnitude or the direction of background magnetic field changes, $\theta$ changes and thus the total polarizability. As a result, $\Delta\alpha(\omega_l)$ deviates from 0, leading to a lower excitation rate and a lattice-induced frequency shift of the transition.

Moreover, we also tune the direction of $\vec{B}_{bias}$ relative to $\vec{k}_{431}$ to maximize the excitation rate. When $\phi$ is close to 132°, we can achieve an excitation rate close to 0.8 and a spectral linewidth below 10 Hz. Figure 1(c) shows one of the spectra observed when the duration of the probe light $T_p$= 100 ms. When we increase the probe time $T_p$= 200 ms, the spectral linewidth gets to 4.3 Hz, as shown in Fig. 1(d). Further increase of $T_p$ does not decrease the spectral linewidth but reduces the excitation rate, mostly due to limited coherence time of the atoms.

Using the Rabi spectrum shown in Fig. 1(c), we stabilized the 431 nm laser to the atomic transition. An interleaved measurement, usually used to assess systematic shifts, shows that we can achieve a fractional frequency instability of $2\times10^{-15}/\sqrt{\tau}$. It enables us to achieve a frequency measurement precision of $1\times10^{-16}$ in a few minutes.

## III. MEASUREMENT OF DIFFERENTIAL STATIC POLARIZABILITY

The BBR shift is largely determined by the differential static polarizability between the ground and excited states [14]. We measure the differential static polarizability $\Delta\alpha_{431}(0)$ between the energy levels associated with the 431 nm transition by detecting the frequency shift induced when a DC voltage is applied to two ring electrodes, as shown in Fig. 1(a). The DC Stark shift of the 431 nm clock transition can be expressed as

$$\Delta\nu_{\text{DC}} = -\frac{1}{2}\Delta\alpha_{431}(0)\left(\vec{E}_{\text{bg}} + \vec{E}_{\text{DC}}\right)^2, \qquad (2)$$

where $\vec{E}_{\text{DC}}$ and $\vec{E}_{\text{bg}}$ are the applied electric field and the background electric field sensed by the atoms, respectively.

To remove the Stark frequency shift arisen from $\vec{E}_{\text{bg}}$, we separately set the voltage of the two electrodes to three settings, (i) $V_1 = V_{\text{DC}}$ and $V_2 = 0$, (ii) $V_1 = 0$ and $V_2 = 0$, (iii) $V_1 = 0$ and $V_2 = V_{\text{DC}}$. We separately measured the frequency difference $\Delta\nu_{\text{DC-1}}$ between (i) and (ii), and then $\Delta\nu_{\text{DC-2}}$ between (ii) and (iii). The average of $\Delta\nu_{\text{DC-1}}$ and $\Delta\nu_{\text{DC-2}}$ is only related to $-\Delta\alpha_{431}(0)|\vec{E}_{\text{DC}}|^2/2$. We use the $^1S_0$–$^3P_0$ clock transition of $^{171}$Yb at 578 nm to calibrate the magnitude of the electric field $\vec{E}_{\text{DC}}$ sensed by the atoms at a certain applied voltage $V_{\text{DC}}$ since the differential static polarizability of the 578 nm clock transition is accurately determined to be $\Delta\alpha_{578}(0) = 36.2612(7)$ kHz/(kV/cm)$^2$ [20].

Figure 2 shows the DC Stark frequency shift of the 431 nm clock transition versus $\vec{E}_{\text{DC}}$. We fit the data with $\Delta\nu_{\text{DC}} = -\Delta\alpha_{431}(0)|\vec{E}_{\text{DC}}|^2/2$, deriving $\Delta\alpha_{431}(0) = -2.10(4)$ kHz/(kV/cm)$^2$. The uncertainty of $\Delta\alpha_{431}(0)$ is dominated by the statistical uncertainty. The measured $\Delta\alpha_{431}(0)$ is 1/2.4 of the calculated value [13], and its magnitude is nearly 1/17 of that of the 578 nm clock transition. The low differential static polarizability of the 431 nm clock transition indicates it is more than an order of magnitude less sensitive to BBR shift, assuming that the contribution of dynamic parameter accounting for frequency dependence of the state polarizabilities across the BBR spectrum is small [13].

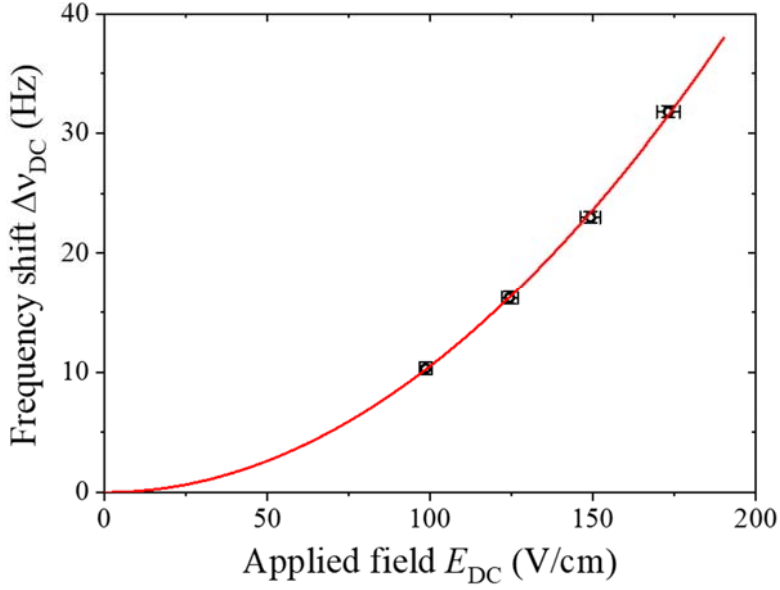

FIG. 2. The DC Stark shift of the 431 nm clock transition versus the applied electric field. The error bars of the applied field and the frequency shift is dominated by the statistical uncertainties. The red solid line is the fitting curve.

## IV. FREQUENCY MEASUREMENT

We investigate the lattice-induced frequency shift of the 431 nm transition by measuring the frequency difference between the transition frequencies under two lattice trap depths, $72E_r$ and $200E_r$ using an interleaved measurement scheme. Figure 3(a) shows a typical lattice-induced frequency shift of the 431 nm transition at different lattice frequencies around 375783 GHz. The error bars are dominated by the statistical uncertainty of the frequency shift, which is larger than those shown in Fig. 2 due to the broadened spectral linewidth. Both the current measurement and the one presented in Fig. 3(b) utilized spectra with a linewidth of 16 Hz. With the aid of a linear fit curve, we determine the lattice-induced Stark shift coefficient is −117(3) μHz/(MHz·$E_r$). When the lattice frequency is kept close to zero-crossing frequency $\nu_0$ within 400 MHz, the lattice-induced frequency shift of the 431 nm transition is 0(4.7) Hz at a lattice trap depth of $100E_r$. We find that when the background magnetic field changes, the zero-crossing lattice frequency $\nu_0$ also changes, with a range of 10 GHz from day to day.

The AC Stark shift induced by the 431 nm probe light is measured at different light intensity. As shown in Fig. 3(b), a linear fit gives a Stark shift coefficient of 0.054(3) Hz/µW. At the operating light power of 45 µW, the probe light shift is estimated to be 2.44(12) Hz. The intensity of the 431 nm probe light is three orders of magnitude larger than that used in the 578 nm transition, leading to more than three orders of magnitude larger shift [22–24].

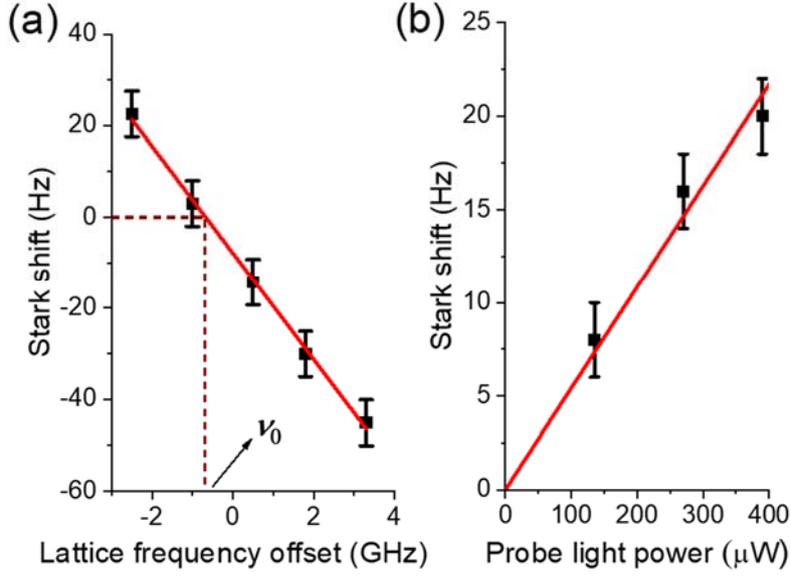

FIG. 3. Measured systematic effects. (a) Lattice Stark shift. The lattice frequency is offset from 375784 GHz. A linear fit (solid line) gives the slope as −11.7(3) Hz/GHz. (b) Probe light shift. A linear fit (solid line) gives the slope as 0.054(3) Hz/µW.

Since here we probe the $^{174}$Yb atoms (bosons) in a one-dimension lattice, $s$-wave scattering may cause a larger collision-induced frequency shift. The collision-induced frequency shift is evaluated by interleaving measurements at different atomic densities. Assuming a same volume, the atomic density is proportional to the number of atoms. The number of atoms is adjusted by changing the duration of the slower light at 399 nm during the first-stage MOT. Using this method, it does not impact the trapping conditions, and thus the density shift is proportional to the number of atoms [23]. We measure a frequency shift of 58(85) mHz when the number of atoms is switched between $3N_{atom}$ and $0.4N_{atom}$, where the number of atoms is $N_{atom} \approx 500$ in normal operation. Assuming the frequency shift is proportional to the atomic density, the typical density shift is estimated to be 22(33) mHz in normal operation.

Other effects include BBR shift, DC Stark shift and Zeeman shift. Based on the measured differential static polarizability $\Delta\alpha_{431}(0)$, we determine that the BBR shift is 0.068(68) Hz, assuming an uncertainty comparable to the BBR shift. The Stark shift due to stray DC electrical field is measured to be far below 1 mHz. To measure the shift due to magnetic field, we increased the bias magnetic field by 1 G. After remeasuring the lattice shift as shown in Fig. 3(a) and adjusting the lattice frequency close to $v_0$, we observed no shift of the 431 nm transition within the measurement uncertainty. Other effects include the uncertainty from the frequency reference $v_{578}$ (see Sec. S3 in Supplemental Material) and the gravitational red shift. The uncertainty of $v_{578}$ is evaluated as 1.3 Hz (accounting for an enlarged BBR shift, which remains negligible compared to the shifts listed in Table I), corresponding to an uncertainty of 1.7 Hz at a wavelength of 431 nm. Therefore, the uncertainty of all the above effects is 1.7 Hz.

We measure the frequency ratio between two optical clocks: clock-431 ($v_{431}$) is stabilized to the $^1S_0$ ($m_F = 0$) $-4f^{13}5d6s^2$ ($J = 2$, $m_F = 0$) transition of $^{174}$Yb, and clock-578 ($v_{578}$) is based on the $^1S_0$–$^3P_0$ transition of $^{171}$Yb. According to the recommended frequency of the $^1S_0$–$^3P_0$ transition of $^{171}$Yb, we determine the frequency of clock-431. Figure 4 shows the measured $v_{431}$ on different days over three months. Each day, we perform lattice shift measurements before and after the frequency ratio measurement to account for potential changes in the background magnetic field. We ensure that the lattice frequency is close to $v_0$ within 400 MHz. The frequency of $v_{431}$ is measured to with a statistical uncertainty of 3.9 Hz (1$\sigma$ standard deviation).

Table I summarizes the systematic frequency shifts and uncertainties for frequency measurement. The total uncertainty is dominated by statistical uncertainty and lattice shift. After correcting frequency shifts listed in Table I, we determine the frequency of the $^1S_0$ ($m_F = 0$) $-4f^{13}5d6s^2$ ($J = 2$, $m_F = 0$) transition in $^{174}$Yb to be 695 175 030 801 776.5(6.3) Hz, corresponding to a frequency ratio of $v_{431}/v_{578}$ =1.341 270 721 706 644(12).

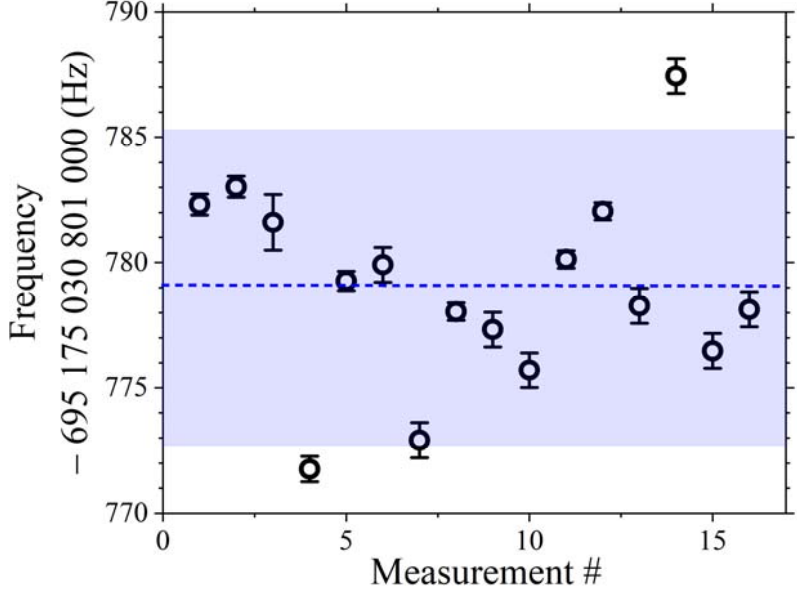

FIG. 4. Measured frequency of the $^1S_0$ ($m_F = 0$) $-4f^{13}5d6s^2$ ($J = 2$, $m_F = 0$) transition of $^{174}$Yb on different days from MJD 60963-61035. The blue shade indicates 1σ standard deviation.

TABLE I. Systematic frequency shifts and uncertainties for frequency measurement of the $^1S_0$ ($m_F = 0$) $-4f^{13}5d6s^2$ ($J = 2$, $m_F = 0$) transition.

|  | Frequency shift (Hz) | Uncertainty (Hz) |
|---|---|---|
| Lattice light | 0 | 4.7 |
| Probe light | 2.44 | 0.12 |
| Density shift | 0.022 | 0.033 |
| Other effects | 0.068 | 1.7 |
| Statistical | 0 | 3.9 |
| Total | 2.5 | 6.3 |

Further reducing the measurement uncertainty of the 431 nm transition frequency requires achieving an even narrower spectral linewidth and maintaining a stable background magnetic field. Variations in the background magnetic field also disrupt atomic coherence, as evidenced by the observed Zeeman shift of the 578 nm clock transition, which indicates field fluctuations on the order of tens of milli-Gauss over a single day. Implementing magnetic shielding or active stabilization techniques is expected to suppress these variations by 2–3 orders of magnitude, thereby enabling more precise studies of this clock transition.

## V. SUMMARY

We report the observation of 4.3 Hz-linewidth spectra for the inner-shell orbital clock transition in $^{174}$Yb at 431 nm. The differential static polarizability of this transition is determined to be −2.10(4) kHz(kV/cm)$^2$, indicating reduced sensitivity to BBR shift. The frequency of the inner-shell orbital clock transition and the frequency ratio between two Yb clocks transitions are measured with uncertainties of $9\times10^{-15}$. These measurements further lay the framework for search for possible variations of the fine-structure constant based on an ensemble of neutral ytterbium atoms.

## VI. ACKNOWLEDGMENTS

This work was supported by the National Natural Science Foundation of China (Grant No. 12334020, 12341404, 12404552) and the National Key R&D Program of China (Grant No. 2022YFB3904001). We are grateful to Yanmei Yu from Institute of Physics, Chinese Academy of Sciences.

# S1. EXPERIMENTAL SETUP FOR LASER FREQUENCY CONTROL AND FREQUENCY RATIO MEASUREMENT

The frequency of the 431 nm probe laser light is referenced to a cavity-stabilized 578 nm laser or further stabilized to the 578 nm clock transition via an optical frequency divider based on an optical frequency comb, as shown in Fig. S1. We use optical fibers to transfer the laser light to the optical frequency divider, reference cavity and the apparatuses for lattice-trapped Yb atoms. The noise introduced during transferring through the optical fibers is cancelled by phase-locking the beating signal between the retroreflected light and local light to a radio frequency (RF) reference, a technique known as fiber noise cancellation (FNC).

The 431 nm laser, which is used for probing the inner-shell orbital clock transition $^1S_0 - 4f^{13}5d6s^2$ ($J = 2$), is the second harmonic of an 862 nm laser. The 862 nm laser is generated by summing the outputs of a diode laser at 1952.8 nm and a fiber laser at 1544.8 nm. The 431 nm laser is purchased from Precilasers. The 578 nm laser light is frequency-stabilized to a 30 cm-long Fabry-Perot cavity. After being stabilized, the 578 nm laser has a linewidth of 0.2 Hz and a fractional frequency instability in the range of $(2\text{-}5) \times 10^{-16}$ at an averaging time of 1 s [S1]. The 578 nm laser light is further frequency-stabilized to the $^1S_0$–$^3P_0$ clock transition by adjusting the RF reference for $FNC_B$. As a result, the frequency of the 578 nm laser light sent to the optical frequency divider is the same as that of the probing the lattice-trapped $^{171}$Yb atoms.

The optical frequency divider sets the frequency ratio between a reference laser at 578 nm and the 862 nm laser as $\nu_{862} = \nu_{578}/R_1$, where $R_1$ is the preset frequency divisor of the optical frequency divider, determined by the settings of direct digital synthesizers (DDSs), including the DDS used as an RF reference for $FNC_A$ and $FNC_3$. The optical frequency comb is frequency-stabilized to a hydrogen maser, while its frequency noise introduced onto the 862 nm laser is eliminated using the transfer oscillator scheme [S2]. Details of optical frequency divider can be found in Ref. [S3]. In brief, we first obtain the beat notes between the 578 nm (862 nm) laser light and their nearby comb teeth. Then, we use mixers and DDSs to remove the carrier-envelope offset frequency ($f_0$) and repetition rate ($f_r$) of the comb from the beat notes. As a result, a virtual beat note between the 578 nm and the 862 nm laser light is obtained, which is free of $f_0$ and $f_r$. This signal is sent to a servo to stabilize the frequency of the 862 nm laser by feedbacking to the current of the 1952.8 nm laser (for fast control) and to a piezoelectric element inside the 1544.8 nm laser (for slow control). The servo bandwidth of the 862 nm laser is approximately 150 kHz. We scan the 431 nm laser for spectrum detection by scanning the RF reference frequency of $FNC_3$. We stabilize the 431 nm laser frequency to the inner-shell orbital clock transition by adjusting the RF reference of $FNC_A$, and the frequency of $FNC_A$ is recorded for calculating the frequency ratio between the two clock transitions.

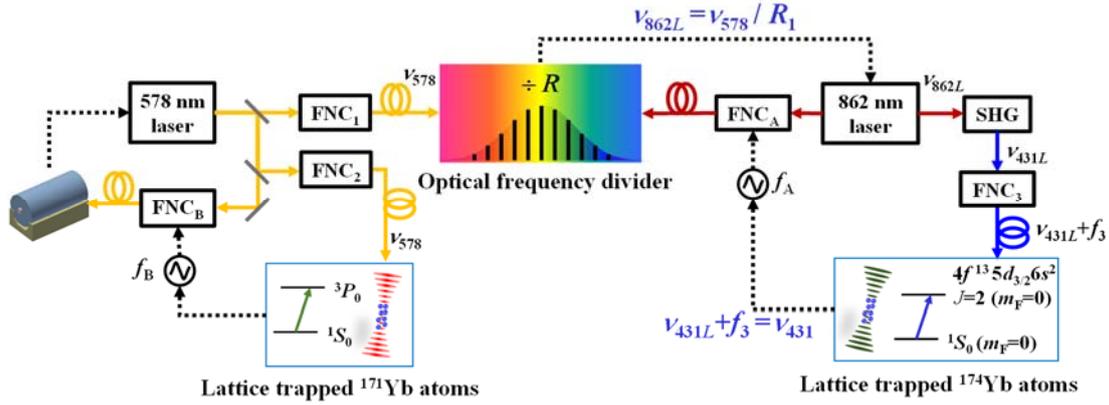

FIG. S1. Experimental setup for laser frequency control and frequency ratio measurement. An optical frequency divider measures the frequency ratio between the 578 nm laser and the fundamental light of the 431 nm laser. Both lasers are frequency-stabilized to the corresponding clock transition in the frequency ratio measurement.

## S2. THE BIAS MAGNETIC FIELD FOR SPECTROSCOPY

To reduce the tensor lattice shift of the 431 nm transition in $^{174}$Yb, three pairs of Helmholtz coils are employed to cancel the static stray magnetic field in three directions and to provide a bias magnetic field. The field strength along each direction is separately calibrated with the $^1S_0$–$^3P_1$ transition. In this context, the $g$ factor of the $^3P_1$ state is $g_J$ ($^3P_1$)=1.49280(4) according to Ref. [S4]. We adjust the current of each pair of coils to control the direction and magnitude of the bias magnetic field. During normal operation, the bias magnetic field has a strength of 1.5 G and forms an angle of nearly 54.8° with respect to the polarization of the lattice light $\vec{e}_{\text{lat}}$.

## S3. THE OPTICAL CLOCK BASED ON THE $^1S_0$–$^3P_0$ TRANSITION OF $^{171}$Yb

Here, we introduce an optical clock based on the $^1S_0$–$^3P_0$ transition of $^{171}$Yb atoms, which is referred to as clock-578 in the main text. The atoms are decelerated and trapped using a two-stage magneto-optical trap (MOT). Then, nearly $10^5$ atoms with a temperature of approximately 15 μK are loaded into a cavity-enhanced optical lattice.

The lattice laser light at 759 nm is generated from a tapered-amplifier laser and transmitted through a piece of polarization-maintenance (PM) optical fiber. At the output of the PM fiber, the light is reflected onto a reflective Bragg grating (RBG) with a spectral bandwidth of 0.05 nm to mitigate the Stark shift induced by background spectra. The diffraction efficiency of the RBG is nearly 90%, depending on the spatial mode of the input light. The power enhancing lattice cavity consists of two highly reflective mirrors with a radius of curvature of 250 mm separated by a distance of 20 cm. These curved mirrors are positioned outside the vacuum chamber, and the cavity is formed vertically to suppress tunneling between adjacent lattice sites. With this cavity configuration, a waist radius of 172 μm in the transverse plane allows for a relatively low atomic density, thereby minimizing the density-induced shift. The cavity is locked to the lattice laser with the Pound-Drever-Hall method by adjusting the voltage applied to a piezoelectric transducer (PZT) attached to one of the mirrors. To minimize the lattice shift, the frequency of the lattice laser is stabilized at the magic wavelength with the aid of an optical frequency comb. About 1000 $^{171}$Yb atoms are trapped when the trap depth is 50 $E_r$.

The atoms are prepared to either of the two nuclear-spin states of $^1S_0$ using a pumping light at 556 nm. Subsequently, we lock the frequency of the 578 nm laser to the $^1S_0$–$^3P_0$ transition of $^{171}$Yb atoms using Rabi spectroscopy. The interleaved measurement of the optical clock demonstrates the frequency instability close to that predicted by a white-noise model, with a value of $1.5 \times 10^{-15}/\sqrt{\tau}$, where $\tau$ represents the averaging time. The value of the absolute frequency of the $^1S_0$–$^3P_0$ transition recommended by the International Committee for Weights and Measures (CIPM) is 518 295 836 590 863.63 (10) Hz [S5]. The systematic uncertainty of clock-578 is evaluated to be 1.3 Hz. This uncertainty is mainly dominated by the black body radiation (BBR) shift, assuming that the temperature uncertainty around the atoms is approximately 300 K. Other frequency shifts are negligible compared to the BBR shift.